\documentclass[11pt,a4paper]{article}
\usepackage{amsfonts}
\usepackage{amsmath,amssymb}
\usepackage{epsfig,graphicx}

\input{tcilatex}

\begin{document}

\begin{center}
{\LARGE \ On Emergent Gauge and Gravity Theories}

\bigskip

\bigskip

\bigskip

\textbf{J.L.~Chkareuli}$^{1,2}$

\bigskip

$^{1}$\textit{Center for Elementary Particle Physics, ITP, Ilia State
University, 0162 Tbilisi, Georgia}

$^{2}$\textit{\textit{A}ndronikashvili Institute of Physics, 0177 Tbilisi,
Georgia\ }

\bigskip

\bigskip

\bigskip \bigskip

\bigskip 

\bigskip
\end{center}

\begin{abstract}
We present some general approach to emergent gauge theories and consider in
significant detail the emergent tensor field gravity case. In essence, an
arbitrary local theory of a symmetric two-tensor field $H_{\mu \nu }$ in
Minkowski spacetime is considered, in which the equations of motion are
required to be compatible with a nonlinear $\sigma $ model type
length-fixing constraint $H_{\mu \nu }^{2}=\pm M^{2}$ leading to spontaneous
Lorentz invariance violation, SLIV ($M$ is the proposed scale for SLIV).
Allowing the parameters in the Lagrangian to be adjusted so as to be
consistent with this constraint, the theory turns out to correspond to
linearized general relativity in the weak field approximation, while some of
the massless tensor Goldstone modes appearing through SLIV are naturally
collected in the physical graviton. The underlying diffeomophism invariance
emerges as a necessary condition for the tensor field $H_{\mu \nu }$ not to
be superfluously restricted in degrees of freedom, apart from the constraint
due to which the true vacuum in the theory is chosen by SLIV. The emergent
theory appears essentially nonlinear, when expressed in terms of the pure
Goldstone tensor modes and contains a plethora of new Lorentz and $CPT$
violating couplings. However, these couplings do not lead to physical
Lorentz violation once this tensor field gravity is properly extended to
conventional general relativity.
\end{abstract}

\bigskip

\bigskip

\begin{center}
\bigskip {\scriptsize \ Invited talk given at the International Workshop
\textquotedblleft Low dimensional physics and gauge
principles\textquotedblright }

{\scriptsize \ 21-29 September 2011, Yerevan-Tbilisi }
\end{center}

\bigskip


\thispagestyle{empty}\newpage

\section{Introduction}

It is conceivable that spontaneous Lorentz invariance violation (SLIV) could
provide a dynamical approach to quantum electrodynamics, gravity and
Yang-Mills theories with photon, graviton and gluons appearing as massless
Nambu-Goldstone bosons\ \cite{bjorken} (for some later developments see \cite%
{cfn,kraus,kos}). However, in contrast to spontaneous violation of internal
symmetries, SLIV seems not to necessarily implies a physical breakdown of
Lorentz invariance. Rather, when appearing in a gauge theory framework, this
may eventually result in noncovariant gauge choice in an otherwise gauge
invariant and Lorentz invariant theory.

Remarkably, a possible source for such a kind of the unobserved SLIV could
provide the nonlinearly realized Lorentz symmetry for underlying vector
field $A_{\mu }$ through its length-fixing constraint
\begin{equation}
A_{\mu }A^{\mu }=n^{2}M^{2}\text{ , \ \ }n^{2}\equiv n_{\nu }n^{\nu }=\pm 1
\label{const}
\end{equation}%
(where $n_{\mu }$ is a properly oriented unit Lorentz vector, while $M$ is
the proposed SLIV scale) rather than some vector field potential. This
constraint in the gauge invariant QED framework was first studied by Nambu
\cite{nambu} a long ago, and in more detail (including the higher order
corrections , extensions to spontaneously broken massive QED and non-Abelian
theories etc.) in the last years \cite{az}. The constraint (\ref{const}),
which in fact is very similar to the constraint appearing in the nonlinear $%
\sigma $-model for pions \cite{GL}, means in essence that the vector field $%
A_{\mu }$ develops some constant background value and the Lorentz symmetry $%
SO(1,3)$ formally breaks down to $SO(3)$ or $SO(1,2)$ depending on the
time-like ($n^{2}>0$) or space-like ($n^{2}<0$) nature of SLIV. The point
is, however, that, in sharp contrast to the nonlinear $\sigma $ model for
pions, the nonlinear QED theory, due to the starting gauge invariance
involved, ensures that all physical Lorentz violating effects are proved to
be strictly cancelled.

Extending the above argumentation, we consider here spontaneous Lorentz
violation realized through a nonlinear length-fixing tensor field constraint
of the type
\begin{equation}
H_{\mu \nu }H^{\mu \nu }=\mathfrak{n}^{2}M^{2}\text{ , \ \ \ \ }\mathfrak{n}%
^{2}\equiv \mathfrak{n}_{\mu \nu }\mathfrak{n}^{\mu \nu }=\pm 1\text{ }
\label{const3}
\end{equation}%
where $\mathfrak{n}_{\mu \nu }$ is a properly oriented `unit' Lorentz
tensor, while $M$ is the scale for Lorentz violation. Such a type of SLIV
implemented into the tensor field gravity theory, which mimics linearized
general relativity in Minkowski space-time, induces massless tensor
Goldstone modes some of which can naturally be collected in the physical
graviton \cite{cjt}. Again, the theory appears essentially nonlinear and
contains a plethora of new Lorentz and $CPT$ violating couplings. However,
these couplings do not lead to physical Lorentz violation once this tensor
field gravity is properly extended to conventional general relativity.

\section{Emergent gauge symmetries}

Speaking still about vector field theories, the most important side of the
nonlinear vector field constraint (\ref{const}) was shown \cite{cj} to be
that one does not need to specially postulate the starting gauge invariance
in the framework of an arbitrary relativistically invariant Lagrangian which
is proposed only to possess some global internal symmetry. Indeed, the SLIV
conjecture (\ref{const}) happens to be powerful enough by itself to require
gauge invariance, provided that we allow the parameters in the corresponding
Lagrangian density to be adjusted so as to ensure self-consistency without
losing too many degrees of freedom. Namely, due to the spontaneous Lorentz
violation determined by the constraint (\ref{const}), the true vacuum in
such a theory is chosen so that this theory acquires on its own a gauge-type
invariance, which gauges the starting global symmetry of the interacting
vector and matter fields involved. In essence, the gauge invariance (with a
proper gauge-fixing term) appears as a necessary condition for these vector
fields not to be superfluously restricted in degrees of freedom.

Let us dwell upon this point in more detail. Generally, while a conventional
variation principle requires the equations of motion to be satisfied, it is
possible to eliminate one component of a general 4-vector field $A_{\mu }$,
in order to describe a pure spin-1 particle by imposing a supplementary
condition. In the massive vector field case there are three physical spin-1
states to be described by the $A_{\mu }$ field. Similarly in the massless
vector field case, although there are only two physical (transverse) photon
spin states, one cannot construct a massless 4-vector field $A_{\mu }$ as a
linear combination of creation and annihilation operators for helicity $\pm
1 $ states in a relativistically covariant way, unless one fictitious state
is added \cite{GLB}. So, in both the massive and massless vector field
cases, only one component of the $A_{\mu }$ field may be eliminated and
still preserve Lorentz invariance. Once the SLIV constraint (\ref{const}) is
imposed, it is therefore not possible to satisfy another supplementary
condition, since this would superfluously restrict the number of degrees of
freedom for the vector field. In fact a further reduction in the number of
independent $A_{\mu }$ components would make it impossible to set the
required initial conditions in the appropriate Cauchy problem and, in
quantum theory, to choose self-consistent equal-time commutation relations
\cite{ogi3}.

We now turn to the question of the consistency of a constraint with the
equations of motion for a general 4-vector field $A_{\mu }$ Actually, there
are only two possible covariant constraints for such a vector field in a
relativistically invariant theory - the holonomic SLIV constraint, $%
C(A)=A_{\mu }A^{\mu }-n^{2}M^{2}=0$ (\ref{const}), and the non-holonomic
one, known as the Lorentz condition, $C(A)=\partial _{\mu }A^{\mu }=0$. In
the presence of the SLIV constraint $C(A)=A^{\mu }A_{\mu }-n^{2}M^{2}=0$, it
follows that the equations of motion can no longer be independent. The
important point is that, in general, the time development would not preserve
the constraint. So the parameters in the Lagrangian have to be chosen in
such a way that effectively we have one less equation of motion for the
vector field. This means that there should be some relationship between all
the (vector and matter) field Eulerians ($E_{A}$, $E_{\psi }$, ...) involved%
\footnote{$E_{A}$ stands for the vector-field Eulerian $(E_{A})^{\mu }\equiv
\partial L/\partial A_{\mu }-\partial _{\nu }[\partial L/\partial (\partial
_{\nu }A_{\mu })].$ We use similar notations for other field Eulerians as
well.}. Such a relationship can quite generally be formulated as a
functional - but by locality just a function - of the Eulerians, $%
F(E_{A},E_{\psi },...)$, being put equal to zero at each spacetime point
with the configuration space restricted by the constraint $C(A)=0$:
\begin{equation}
F(C=0;\text{ \ }E_{A},E_{\psi },...)=0\text{ .}  \label{FF}
\end{equation}%
This relationship must satisfy the same symmetry requirements of Lorentz and
translational invariance, as well as all the global internal symmetry
requirements, as the general starting Lagrangian $L(A,\psi ,...)$ does. We
shall use this relationship in subsequent sections as the basis for gauge
symmetry generation in the SLIV constrained vector and tensor field theories.

Let us now consider a \textquotedblleft Taylor expansion" of the function $F$
expressed as a linear combination of terms involving various field
combinations multiplying or derivatives acting on the Eulerians\footnote{%
The Eulerians are of course just particular field combinations themselves
and so this \textquotedblleft expansion" at first includes higher powers and
higher derivatives of the Eulerians.}. The constant term in this expansion
is of course zero since the relation (\ref{FF}) must be trivially satisfied
when all the Eulerians vanish, i.e. when the equations of motion are
satisfied. We now consider just the terms containing field combinations (and
derivatives) with mass dimension 4, corresponding to the Lorentz invariant
expressions
\begin{equation}
\partial _{\mu }(E_{A})^{\mu },\text{ }A_{\mu }(E_{A})^{\mu },\text{ }%
E_{\psi }\psi ,\text{ }\overline{\psi }E_{\overline{\psi }}.  \label{fff}
\end{equation}%
All the other terms in the expansion contain field combinations and
derivatives with higher mass dimension and must therefore have coefficients
with an inverse mass dimension. We expect the mass scale associated with
these coefficients should correspond to a large fundamental mass (e.g. the
Planck mass $M_{P}$). Hence we conclude that such higher dimensional terms
must be highly suppressed and can be neglected. A priori these neglected
terms could lead to the breaking of the spontaneously generated gauge
symmetry at high energy. However it could well be that a more detailed
analysis would reveal that the imposed SLIV constraint requires an exact
gauge symmetry. Indeed, if one uses classical equations of motion, a gauge
breaking term will typically predict the development of the
\textquotedblleft gauge\textquotedblright\ in a way that is inconsistent
with our gauge fixing constraint $C(A)=0$. Thus the theory will generically
only be consistent if it has exact gauge symmetry\footnote{%
The other possible Lorentz covariant constraint $\partial _{\mu }A^{\mu }=0,$
while also being sensitive to the form of the constraint-compatible
Lagrangian, leads to massive QED and massive Yang-Mills theories \cite{ogi3}.%
}.

\section{Deriving diffeomorphism invariance\textbf{\ }}

We now illustrate these ideas by the example of the emergent tensor field
gravity case. Let us consider an arbitrary relativistically invariant
Lagrangian $\mathcal{L}(H_{\mu \nu },\phi )$ of one symmetric two-tensor
field $H_{\mu \nu }$ and one real scalar field $\phi $ as the simplest
possible matter in the theory taken in Minkowski spacetime. We restrict
ourselves to the minimal interactions. In contrast to vector fields, whose
basic interactions contain dimensionless coupling constants, for tensor
fields the interactions with coupling constants of dimensionality $sm^{1}$
are essential. We first turn to the possible supplementary conditions which
can be imposed on the tensor fields $H_{\mu \nu }$ in the Lagrangian $%
\mathcal{L}$, possessing still only a global Lorentz (and translational)
invariance, in order to finally establish its form. The SLIV constraint (\ref%
{const3}), as it usually is when considering a system with holonomic
constraints, can equivalently be presented in terms of some Lagrange
multiplier term in the properly extended Lagrangian $\mathcal{L}^{\prime
}(H_{\mu \nu },\phi ,\mathcal{\lambda })$ rather than to be substituted into
the starting one $\mathcal{L}(H_{\mu \nu },\phi )$ prior to the variation of
the action. Writing $\mathcal{L}^{\prime }(H_{\mu \nu },\phi ,\mathcal{%
\lambda })$ as%
\begin{equation}
\mathcal{L}^{\prime }(H_{\mu \nu },\phi ,\mathcal{\lambda })=\mathcal{L}%
(H_{\mu \nu },\phi )-\frac{1}{4}\mathfrak{\lambda }\left( H_{\mu \nu }H^{\mu
\nu }-\mathfrak{n}^{2}M^{2}\right) ^{2}  \label{lag22}
\end{equation}%
and varying with respect to the auxiliary field $\mathcal{\lambda }(x)$ one
has just the SLIV condition (\ref{const3}). \ Our choice of the quadratic
form of the Lagrange-multiplier term \cite{kkk} is only related to the fact
that the equations of motion for $H_{\mu \nu }$ in this case are independent
of the $\mathcal{\lambda }(x)$ which entirely decouples from them rather
than acts as some extra source of energy-momentum density, as it would be
for the linear Lagrange multiplier term that could make the subsequent
consideration to be more complicated. So, as soon as the constraint (\ref%
{const3}) holds%
\begin{equation}
\mathcal{C}(H_{\mu \nu })=H_{\mu \nu }H^{\mu \nu }-\mathfrak{n}^{2}M^{2}=0
\label{eqss1}
\end{equation}%
one has the equations of motion for $H_{\mu \nu }$ expressed through its
Eulerian $(\mathcal{E}_{H})^{\mu \nu }$
\begin{equation}
(\mathcal{E}_{H})^{\mu \nu }\equiv \partial \mathcal{L}/\partial H_{\mu \nu
}-\partial _{\rho }[\partial \mathcal{L}/\partial (\partial _{\rho }H_{\mu
\nu })]=0\text{ }  \label{eqss}
\end{equation}%
which is determined solely by the starting Lagrangian $\mathcal{L}(H_{\mu
\nu },\phi ).$

Despite the SLIV constraint (\ref{const3}) the tensor field $H_{\mu \nu }$,
both massive and massless, still contains many superfluous components which
are usually eliminated by imposing some supplementary conditions. In the
massive tensor field case there are five physical spin-$2$ states to be
described by $H_{\mu \nu }$. Similarly, in the massless tensor field case,
though there are only two physical (transverse) spin states associated with
graviton, one cannot construct a symmetric two-tensor field $H_{\mu \nu }$
as a linear combination of creation and annihilation operators for helicity $%
\pm 2$ states unless three (and $2j-1,$ in general, for the spin $j$
massless field) fictitious states with other helicities are added \cite%
{GL,GLB}. So, in both massive and massless tensor field cases only five
components in the $10$-component tensor field $H_{\mu \nu }$ may be at most
eliminated \ so as to preserve the Lorentz invariance. However, once the
SLIV constraint (\ref{eqss1}) is already imposed, four extra supplementary
conditions are only possible. Normally they should exclude the spin $1$
states which are still left in the theory\footnote{%
These spin $1$ states must necessarily be excluded as the sign of the energy
for spin $1$ is always opposite to that for spin $2$ and $0$} and are
described by some of the components of the tensor $H_{\mu \nu }$. Usually,
they (and one of the spin $0$ states) are excluded by the conventional
harmonic gauge condition
\begin{equation}
\partial ^{\mu }H_{\mu \nu }-\partial _{\nu }H_{tr}/2=0\text{ }.  \label{HL}
\end{equation}%
or some of its analogs (see section 4). In fact, there should not be more
supplementary conditions - otherwise, this would superfluously restrict the
number of degrees of freedom for the spin $2$ tensor field which is
inadmissible.

Under this assumption of not getting too many constraints, we shall now
derive gauge invariance of the Lagrangian $\mathcal{L}(H_{\mu \nu },\phi )$.
Actually, we turn to the question of the consistency of the SLIV constraint
with the equations of motion for a general symmetric tensor field $H_{\mu
\nu }$. For an arbitrary Lagrangian $\mathcal{L}(H_{\mu \nu },\phi )$, the
time development of the fields would not preserve the constraint (\ref{eqss1}%
). So the parameters in the Lagrangian must be chosen so as to give a
relationship between the Eulerians for the tensor and matter fields of the
type
\begin{equation}
\mathcal{F}^{\mu }(\mathcal{C}=0;\text{ \ }\mathcal{E}_{H},\mathcal{E}_{\phi
},...)=0\text{ \ \ \ }(\mu =0,1,2,3).  \label{fmu}
\end{equation}%
which, in contrast to the relationship (\ref{FF}), transforms in general as
a Lorentz vector in the tensor field case. As a result, four additional
equations for the tensor field $H_{\mu \nu }$ which appear by taking
4-divergence of the tensor field equations of motion (\ref{eqss})
\begin{equation}
\partial _{\mu }(\mathcal{E}_{H})^{\mu \nu }=0\   \label{1111}
\end{equation}%
will not produce supplementary conditions at all once the SLIV condition (%
\ref{eqss1}) occurs. In fact, due to the relationship (\ref{fmu}) these
equations (\ref{1111}) are satisfied identically or as a result of the
equations of motion of all the fields involved. This implies that in the
absence of the equations of motion there must hold a general off-shell
identity of the type%
\begin{equation}
\partial _{\mu }(\mathcal{E}_{H})^{\mu \nu }=P_{\alpha \beta }^{\nu }(%
\mathcal{E}_{H})^{\alpha \beta }+Q^{\nu }\mathcal{E}_{\phi }  \label{id1}
\end{equation}%
where $P_{\alpha \beta }^{\nu }$ and $Q^{\nu }$ are some operators acting on
corresponding Eulerians of tensor and scalar fields (for this form of $%
\partial _{\mu }(\mathcal{E}_{H})^{\mu \nu }$ the second equation in (\ref%
{1111}) is trivially satisfied). The simplest conceivable forms of these
operators are
\begin{eqnarray}
P_{\alpha \beta }^{\nu } &=&p_{1}\eta ^{\nu \rho }(H_{\alpha \rho }\partial
_{\beta }+H_{\rho \beta }\partial _{\alpha }+\partial _{\rho }H_{\alpha
\beta })\text{ , \ }  \label{pp} \\
Q^{\nu } &=&q_{1}\eta ^{\nu \rho }\partial _{\rho }\phi \text{ \ }  \nonumber
\end{eqnarray}%
in which only terms with constants $p_{1}$ and $q_{1}$ of dimensionality $%
cm^{1}$ appear essential. This identity (\ref{id1}) implies then the
invariance of $\mathcal{L}(H_{\mu \nu },\phi )$ under the local
transformations of tensor and scalar fields whose infinitesimal form is
given by
\begin{eqnarray}
\delta H_{\mu \nu } &=&\partial _{\mu }\xi _{\nu }+\partial _{\nu }\xi _{\mu
}  \label{sc1} \\
&&+p_{1}(\partial _{\mu }\xi ^{\rho }H_{\rho \nu }+\partial _{\nu }\xi
^{\rho }H_{\mu \rho }+\xi ^{\rho }\partial _{\rho }H_{\mu \nu }),\text{ }
\nonumber \\
\text{\ }\delta \phi &=&q_{1}\xi ^{\rho }\partial _{\rho }\phi  \nonumber
\end{eqnarray}%
where $\xi ^{\mu }(x)$ is an arbitrary 4-vector function, only being
required to conform with the nonlinear constraint (\ref{const3}).
Conversely, the identity (\ref{id1}) in its turn follows from the invariance
of the Lagrangian $\mathcal{L}(H_{\mu \nu },\phi )$ under the
transformations (\ref{sc1}). Both direct and converse assertions are in fact
particular cases of Noether's second theorem \cite{noeth}.

An important point is that the operators $P_{\alpha \beta }^{\nu }$ and $%
Q^{\nu }$ (\ref{pp}) were chosen in a way that the corresponding
transformations (\ref{sc1}) could generally constitute a group (that is the
Lie structure relation holds). This is why all three terms in the symmetric
operator $P_{\alpha \beta }^{\nu }$ (\ref{pp}) are taken with the same
constant, though the third term in it might enter with some different
constant. Remarkably, though the transformations (\ref{sc1}) were only
restricted to form a group, this emergent symmetry group is proved, as one
can readily confirm, to be nothing but the diff invariance. Indeed, for the
quantity%
\begin{equation}
g_{\mu \nu }=\eta _{\mu \nu }+p_{1}H_{\mu \nu }  \label{metr}
\end{equation}%
the tensor field transformation (\ref{sc1}) may be written in a form%
\begin{equation}
\delta g_{\mu \nu }=p_{1}(\partial _{\mu }\xi ^{\rho }g_{\rho \nu }+\partial
_{\nu }\xi ^{\rho }g_{\mu \rho }+\xi ^{\rho }\partial _{\rho }g_{\mu \nu })
\label{mt}
\end{equation}%
which shows that $g_{\mu \nu }$ transform as the metric tensors in the
Riemannian geometry (the constant $p_{1}$ may be included into the
transformation 4-vector parameter $\xi ^{\mu }(x)$) with general coordinate
transformations, $\delta x^{\mu }=\xi ^{\mu }(x)$. So, we have shown that
the imposition of the SLIV constraint (\ref{const3}) supplements the
starting global Poincare symmetry with the local diff invariance. Otherwise,
the theory would superfluously restrict the number of degrees of freedom for
the tensor field $H_{\mu \nu }$, which would certainly not be allowed.

This SLIV induced gauge symmetry (\ref{sc1}) completely determines now the
Lagrangian $\mathcal{L}(H_{\mu \nu },\phi )$. Indeed, in the weak field
approximation (when $\delta H_{\mu \nu }=\partial _{\mu }\xi _{\nu
}+\partial _{\nu }\xi _{\mu }$) this symmetry gives the well-known
linearized gravity Lagrangian
\begin{equation}
\mathcal{L}(H_{\mu \nu },\phi )=\mathcal{L}(H)+\mathcal{L}(\phi )+\mathcal{L}%
_{int}  \label{tl}
\end{equation}%
consisted of the $H$ field kinetic \ term of the form%
\begin{eqnarray}
\mathcal{L}(H) &=&\frac{1}{2}\partial _{\lambda }H^{\mu \nu }\partial
^{\lambda }H_{\mu \nu }-\frac{1}{2}\partial _{\lambda }H_{tr}\partial
^{\lambda }H_{tr}  \label{fp} \\
&&-\partial _{\lambda }H^{\lambda \nu }\partial ^{\mu }H_{\mu \nu }+\partial
^{\nu }H_{tr}\partial ^{\mu }H_{\mu \nu }\text{ ,}  \nonumber
\end{eqnarray}%
($H_{tr}$ stands for a trace of the $H_{\mu \nu },$ $H_{tr}=\eta ^{\mu \nu
}h_{\mu \nu }$), and the free scalar field Lagrangian $\mathcal{L}(\phi )$
and interaction term\ $\mathcal{L}_{int}=(1/M_{P})H_{\mu \nu }T^{\mu \nu
}(\phi ),$ where $T^{\mu \nu }(\phi )$ is a conventional energy-momentum
tensor for scalar field. Besides, the proportionality coefficient $p_{1}$ in
the metric (\ref{metr}) was chosen to be inverse just to the Planck mass $%
M_{P}$. It is clear that, in contrast to the free field terms given above by
$\mathcal{L}(H)$ and $\mathcal{L}(\phi )$, the interaction term $\mathcal{L}%
_{int}$ is only approximately invariant under the diff transformations in
the weak field limit.

To determine a complete theory, one should consider the full variation of
the Lagrangian $\mathcal{L}$ as function of metric $g_{\mu \nu }$ and its
derivatives (including the second order ones), and solve a general identity
of the type%
\begin{equation}
\delta \mathcal{L}(g_{\mu \nu },g_{\mu \nu ,\lambda },g_{\mu \nu ,\lambda
\rho };\phi ,\phi _{,\lambda })=\partial _{\mu }X^{\mu }  \label{iden}
\end{equation}%
(subscripts after commas denote derivatives) which contains an unknown
vector function $X^{\mu }$. The latter must be constructed from the fields
and local transformation parameters $\xi ^{\mu }(x)$ taking into account the
requirement of compatibility with the invariance of the $\mathcal{L}$ under
transformations of the Lorentz group and translations. Following this
procedure \cite{ogi4} for the metric and scalar field variations (\ref{mt}, %
\ref{sc1}) conditioned by SLIV constraint (\ref{const3}), one can eventually
find the total Lagrangian $\mathcal{L}$ which \ is turned out to be properly
expressed in terms of quantities similar to the basic ones in the Riemannian
geometry (like as metric, connection, curvature etc.). Actually, this theory
successfully mimics general relativity that allows us to conclude that the
Einstein equations could be really derived in the flat Minkowski spacetime
provided that Lorentz symmetry in it is spontaneously broken.

\section{Graviton as a Goldstone boson\textbf{\ }}

Let us turn now to spontaneous Lorentz violation in itself which is caused
by the nonlinear tensor field constraint (\ref{const3}). This constraint
means in essence that the tensor field $H_{\mu \nu }$ develops the vev
configuration
\begin{equation}
<H_{\mu \nu }(x)>\text{ }=\mathfrak{n}_{\mu \nu }M  \label{v}
\end{equation}%
determined by the matrix $\mathfrak{n}_{\mu \nu }$, and starting Lorentz
symmetry $SO(1,3)$ of the Lagrangian $\mathcal{L}(H_{\mu \nu },\phi )$ given
in (\ref{tl}) formally breaks down at a scale $M$ to one of its subgroup
thus producing a corresponding number of the Goldstone modes. In this
connection the question about other components of a symmetric two-index
tensor $H_{\mu \nu },$ aside from the pure Goldstone ones, naturally arises.
Remarkably, they are turned out to be the pseudo-Goldstone modes (PGMs) in
the theory. Indeed, although we only propose the Lorentz invariance of the
Lagrangian $\mathcal{L}(H_{\mu \nu },\phi )$, the SLIV constraint (\ref%
{const3}) possesses formally a much higher accidental symmetry. This is in
fact $SO(7,3)$ symmetry of the length-fixing bilinear form (\ref{const3}).
This symmetry is spontaneously broken side by side with Lorentz symmetry at
scale $M.$ Assuming a minimal vacuum configuration in the $SO(7,3)$ \ space
with the vevs (\ref{v}) developed on only one $H_{\mu \nu }$ component, we
have the time-like ($SO(7,3)$ $\rightarrow SO(6,3)$) or space-like ($SO(7,3)$
$\rightarrow SO(7,2)$) violations of the accidental symmetry depending on
the sign of $\mathfrak{n}^{2}=\pm 1$ in (\ref{const3}), respectively.
According to the number of the broken generators just nine massless NG modes
appear in both of cases. Together with an effective Higgs component, on
which the vev is developed, they complete the whole ten-component symmetric
tensor field $H_{\mu \nu }$ of our Lorentz group. Some of them are true
Goldstone modes of spontaneous Lorentz violation, the others are PGMs since,
as was mentioned, an accidental $SO(7,3)$ is not shared by the whole
Lagrangian $\mathcal{L}(H_{\mu \nu },\phi )$ given in (\ref{tl}). Notably,
in contrast to the known scalar PGM case \cite{GL}, they remain strictly
massless being protected by the simultaneously generated diff invariance.

Now, one can rewrite the Lagrangian $\mathcal{L}(H_{\mu \nu },\phi )$ in
terms of the Goldstone modes explicitly using the SLIV constraint (\ref%
{const3}). For this purpose let us take the following handy parameterization
for the tensor field $H_{\mu \nu }$ in the Lagrangian $\mathcal{L}(H_{\mu
\nu },\phi )$:
\begin{equation}
H_{\mu \nu }=h_{\mu \nu }+\frac{n_{\mu \nu }}{n^{2}}(\mathfrak{n}_{\alpha
\beta }H^{\alpha \beta })\text{ , \ \ \ }\mathfrak{n}_{\mu \nu }h^{\mu \nu
}=0  \label{par}
\end{equation}%
where $h_{\mu \nu }$ corresponds to the pure Goldstonic modes, while the
effective \textquotedblleft Higgs" mode (or the $H_{\mu \nu }$ component in
the vacuum direction) is
\begin{equation}
\text{\ }\mathfrak{n}_{\alpha \beta }H^{\alpha \beta }\text{\ }=(M^{2}-%
\mathfrak{n}^{2}h^{2})^{\frac{1}{2}}=M-\frac{\mathfrak{n}^{2}h^{2}}{2M}%
+O(1/M^{2})  \label{constr1}
\end{equation}%
taking, for definiteness, the positive sign for the square root and
expanding it in powers of $h^{2}/M^{2}$, $h^{2}\equiv h_{\mu \nu }h^{\mu \nu
}$. Putting then the parameterization (\ref{par}) with the SLIV constraint (%
\ref{constr1}) into Lagrangian $\mathcal{L}(H_{\mu \nu },\phi )$ given in (%
\ref{tl}), one readily comes to the truly Goldstonic tensor field gravity
Lagrangian $\mathcal{L}(h_{\mu \nu },\phi )$ containing infinite series in
powers of the $h_{\mu \nu }$ modes, which we will not display here due to
its excessive length (see \cite{cjt}).

Together with the Lagrangian $\mathcal{L}(h_{\mu \nu },\phi )$ one must be
also certain about the gauge fixing terms, apart from a general Goldstonic \
"gauge" \ $\mathfrak{n}_{\mu \nu }h^{\mu \nu }=0$ given above (\ref{par}).
Remarkably, the simplest set of conditions being compatible with the latter
is turned out to be
\begin{equation}
\partial ^{\rho }(\partial _{\mu }h_{\nu \rho }-\partial _{\nu }h_{\mu \rho
})=0  \label{gauge}
\end{equation}%
(rather than harmonic gauge conditions (\ref{HL})) which also automatically
eliminates the (negative-energy) spin $1$ states in the theory. So, with the
Lagrangian $\mathcal{L}(h_{\mu \nu },\phi )$ and the supplementary
conditions\ (\ref{par}) and (\ref{gauge}) lumped together, one eventually
comes to the working model for the Goldstonic tensor field gravity.
Generally, from ten components in the symmetric-two $h_{\mu \nu }$ tensor,
four components are excluded by the supplementary conditions (\ref{par}) and
(\ref{gauge}). For a plane gravitational wave propagating, say, in the $z$
direction another four components can also be eliminated. This is due to the
fact that the above supplementary conditions still leave freedom in the
choice of a coordinate system, $x^{\mu }\rightarrow $ $x^{\mu }-\xi ^{\mu
}(t-z/c),$ much as takes place in standard GR. Depending on the form of the
vev tensor $\mathfrak{n}_{\mu \nu }$, the two remaining transverse modes of
the physical graviton may consist solely of Lorentz Goldstone modes or of
Pseudo Goldstone modes or include both of them.

\section{Summary\textbf{\ }and outlook}

We presented here some approach to emergent gauge theories and considered in
significant detail the emergent tensor field gravity case. Our main result
can be summarized in a form of a general \textit{Emergent Gauge Symmetry}
\textit{\ (EGS) conjecture}:

\textit{Let there be given an interacting field system }$\{\boldsymbol{A}%
_{\mu },...,\boldsymbol{\phi },$ $\boldsymbol{\psi },$ $H_{\mu \nu }\}$%
\textit{\ containing some vector field (or vector field multiplet) }$%
\boldsymbol{A}_{\mu }$\textit{\ or/and tensor field }$H_{\mu \nu }$ \textit{%
in an arbitrary Lorentz invariant Lagrangian of scalar (}$\boldsymbol{\phi }$%
), \textit{fermion (}$\boldsymbol{\psi }$\textit{) and other matter field
multiplets, which possesses only global Abelian or non-Abelian symmetry }$G$
\textit{(and only conventional global Lorentz invariance in a pure tensor
field case). Suppose that one of fields in a given field system is subject
to the nonlinear }$\sigma $ \textit{model type} \textit{"length-fixing"
constraint, say, }$\boldsymbol{A}_{\mu }^{2}=M^{2}$ \textit{(for vector
fields) or} $\boldsymbol{\phi }^{2}=M^{2}$ \textit{(for scalar fields) or }$%
H_{\mu \nu }^{2}=M^{2}$ \textit{(for tensor field)}.\textit{\ Then, since
the time development would not in general preserve this constraint, the
parameters in their common Lagrangian }$L(\boldsymbol{A}_{\mu },...,%
\boldsymbol{\phi },$ $\boldsymbol{\psi };$ $H_{\mu \nu })$ \textit{will
adjust themselves in such a way that effectively we have less independent
equations of motion for the field system taken. This means that there should
be some relationship between Eulerians of all the fields involved to which
Noether's second theorem can be applied. As a result, one comes to the
conversion of the global symmetry }$G$\textit{\ into the local symmetry }$%
G_{loc}$, \textit{being exact or spontaneously broken depending on whether
vector or scalar fields are constrained, and to the conversion of global
Lorentz invariance into diffeomorphism invariance for the constrained tensor
field. }

Applying the EGS conjecture to tensor field theory case we found that the
only possible local theory of a symmetric two-tensor field $H_{\mu \nu }$ in
Minkowski spacetime which is compatible with SLIV constraint $H_{\mu \nu
}^{2}=\pm M^{2}$ is turned out to be linearized general relativity in the
weak field approximation. When expressed in terms of the pure tensor
Goldstone modes this theory is essentially nonlinear and contains a variety
of \ Lorentz and $CPT$ violating couplings. Nonetheless, as was shown in the
recent calculations \cite{cjt}, all the SLIV effects turn out to be strictly
cancelled in the lowest order gravity processes as soon as the tensor field
gravity theory is properly extended to general relativity. So, the nonlinear
SLIV condition \ being applied both in vector and tensor field theories, due
to which true vacuum is chosen and Goldstonic gauge fields are generated,
may provide a dynamical setting for all underlying internal and spacetime
local symmetries involved. However, this gauge theory framework, uniquely
emerging for the length-fixed vector and tensor fields, makes in turn this
SLIV to be physically unobservable.

From this standpoint, the only way for physical Lorentz violation to appear
would be if the above local invariance were slightly broken at very small
distances controlled by quantum gravity \cite{ck}. The latter could in
general hinder the setting of the required initial conditions in the
appropriate Cauchy problem thus admitting a superfluous restriction of
vector and tensor fields in degrees of freedom through some high-order
operators stemming from the quantum gravity influenced area. This may be a
place where the emergent vector and tensor field theories may drastically
differ from conventional gauge theories that could have some observational
evidence at low energies.

\section*{Acknowledgments}

This paper is largely based on the recent works \cite{cjt,ck,cfn2} carried
out in collaboration with Colin Froggatt, Juansher Jejelava, Zurab
Kepuladze, Holger Nielsen and Giorgi Tatishvili. I would like to thank Oleg
Kancheli and Archil Kobakhidze for interesting discussions and useful
remarks.

\end{document}